*Article*

# From PDF to RAG-Ready: Evaluating Document Conversion Frameworks for Domain-Specific Question Answering

**José Guilherme Marques dos Santos [1,†], Ricardo Yang [1,†], Rui Humberto Pereira [2,3], Alexandre Sousa [3,4], Brígida Mónica Faria [3,5], Henrique Lopes-Cardoso [1,3], José Duarte [3,4], José Luís Reis [2,3], Luís Paulo Reis [1,3], Pedro Pimenta [3,6] and José Paulo Marques dos Santos [2,3,*]**

[1] Faculty of Engineering, University of Porto, 4200-465 Porto, Portugal; up202208081@edu.fc.up.pt (J.G.M.d.S.); up202208465@edu.fe.up.pt (R.Y.); hlc@fe.up.pt (H.L.-C.); lpreis@fe.up.pt (L.P.R.)
[2] Department of Business Administration, University of Maia, 4475-690 Maia, Portugal; rhpereira@umaia.pt (R.H.P.); jreis@umaia.pt (J.L.R.)
[3] LIACC—Artificial Intelligence and Computer Science Laboratory, University of Porto, 4200-465 Porto, Portugal; asousa@umaia.pt (A.S.); monica.faria@ess.ipp.pt (B.M.F.); d012155@umaia.pt (J.D.); ppimenta@ipmaia.pt (P.P.)
[4] Department of Communication Sciences and Information Technologies, University of Maia, 4475-690 Maia, Portugal
[5] School of Health, Polytechnic of Porto, 4200-072 Porto, Portugal
[6] School of Technology and Management, Polytechnic Institute of Maia, 4475-690 Maia, Portugal
[*] Correspondence: jpsantos@umaia.pt
[†] These authors contributed equally to this work.

### Featured Application

A software platform for answering domain-specific queries over collections of PDF documents using Retrieval-Augmented Generation (RAG), applicable to administrative, legal, and regulatory document management in organizations handling sensitive or non-English documentation.

### Abstract

Retrieval-Augmented Generation (RAG) systems depend critically on the quality of document preprocessing, yet no prior study has evaluated PDF processing frameworks by their impact on downstream question-answering accuracy. We address this gap through a systematic comparison of four open-source PDF-to-Markdown conversion frameworks, Docling, MinerU, Marker, and DeepSeek OCR, across 21 pipeline configurations, varying the conversion tool, cleaning transformations, splitting strategy, and metadata enrichment. Evaluation was performed using a 50-question benchmark over a corpus of 36 Portuguese administrative documents (1706 pages, ~492K words), with LLM-as-judge scoring over 50 independent runs per configuration. Statistical significance was assessed via Wilcoxon signed-rank tests with Cohen's d effect sizes. Two baselines bounded the results: naïve PDFLoader (86.2%) and manually curated Markdown (91.3%). Docling with hierarchical splitting and image descriptions achieved the highest automated accuracy (94.1 ± 1.6%), surpassing even manual curation. A per-question-type analysis revealed that table-dependent questions drive the largest accuracy differences, with a 33-percentage-point gap between basic and hierarchical splitting. Metadata enrichment and hierarchy-aware chunking contributed more to accuracy than the conversion framework alone. An exploratory GraphRAG implementation underperformed basic RAG (82% vs. 94.1%). These findings demonstrate that data preparation quality is the dominant factor in RAG system performance.

## 1. Introduction

Retrieval-Augmented Generation (RAG) has emerged in recent times as the dominant paradigm for grounding Large Language Models (LLMs) in domain-specific knowledge. RAG systems contribute to addressing critical limitations such as hallucinations in LLMs' responses, knowledge cutoffs, especially when LLMs are required to deal with domain-specific questions that require precise expertise, and importantly, for the sake of transparency in LLMs' computations, generations, and outputs, the lack of traceability [1,2]. By dynamically retrieving relevant document excerpts at inference time, RAG systems enable LLMs to generate responses anchored in verifiable source material, which improves factual accuracy for knowledge-intensive tasks [1]. Recent studies have been identifying a swiftly expanding landscape of RAG architectures, evaluation frameworks, and application domains [2,3]. Such a dynamic reflects both the evolutionary pace of the technique, despite already showing maturity, and on the other hand, the pressure to deliver reliable, domain-adapted AI systems.

A prevalent approach involves concentrating on the retrieval mechanism, the selection of the embedding model, or the LLM itself when developing a Retrieval-Augmented Generation (RAG) system for answering questions over a corpus of PDF documents. Nevertheless, minimal consideration is given to an earlier and arguably more critical step: the conversion of PDFs into text. The experience detailed herein pertains to the development of such a system for the Personnel Command of the Portuguese Army. Human resources documents necessitated addressing this issue; they yielded, however, some unforeseen outcomes.

RAG [1] operates through a sequence of steps: initially retrieving relevant passages from an external document collection and subsequently feeding them to a large language model (LLM) alongside the user's query. This method ensures that the model's response is anchored in actual source material rather than relying solely on its parametric knowledge derived from transformer pre-training. It is this architecture that has been shown to be effective in mitigating hallucinations and, moreover, to enhance factual accuracy across various knowledge-intensive applications [2,3]. Nonetheless, most RAG studies assume ideal input text, provide insufficient detail regarding preprocessing stages at the outset of the pipeline, or lack transparency concerning the efficiency of this stage, thereby creating a gap that warrants further investigation [2,4,5].

However, the success of any RAG-based system is fundamentally constrained by the quality of the data it retrieves from. If errors are introduced during document preprocessing, such as misread tables, lost document hierarchy, or distorted characters or diacritics, they will propagate directly from the source into the retrieval and generation stages, degrading the accuracy and reliability of the system's outputs [4,6]. Despite this hindrance, which may have impactful consequences for the outcomes, even catastrophic (e.g., inducing hallucinations), the document preprocessing stage remains comparatively understudied. The majority of research effort has been directed at improving retrieval algorithms, reranking strategies, chunking methods (i.e., methods by which documents are divided into smaller text segments for embedding and retrieval), and generation quality [2,3], while the upstream conversion of raw documents into machine-readable formats is often treated as a solved problem or a mere engineering detail.

This gap is particularly consequential when dealing with PDF documents, which remain the most common format for regulatory, legal, administrative, and technical documentation worldwide. PDFs, nonetheless, were created with format-preserving document sharing and printing as the purpose, not as a vehicle for text for LLMs. Therefore, it is with no surprise that it is considered a notoriously difficult protocol to work with programmatically. Unlike HTML or Markdown, the PDF format is focused on the visual aspect and less on the structural ones, such as text sequence, paragraphs, and hierarchical structure. Having format-preserving as a reference, it encodes where characters and graphical elements should appear on a printed page [7,8]. Extracting structured, semantically meaningful text from PDFs, which involves the preservation of section headings, table structures, reading order, and embedded content, is an active research problem with no fully solved solution [9]. The difficulty increases substantially when the document contains special layouts, for example, scanned images, merged table cells, multi-column layouts, form fields, or mathematical formulae and notation. Another source of complexity and errors is special characters and diacritics, common in many languages other than English. For example, the letter in Portuguese "ç" is often misrecognized across multiple conversion tools, an error that directly corrupts retrieval results, which leads to additional pitfalls. The word "caça" (connected with hunting, as in "carne de caça", which means "game meat", or "caça selvagem", which means "wild game") may be recoded to "caca", which literally means feces. Straightforwardly, everyone can anticipate the complete change in meaning that such an erroneous recoding could cause in a sentence.

Several open-source frameworks offer PDF-to-Markdown or PDF-to-structured-text conversion. Among the most common there are:

- Docling [7,8], which uses modular specialized models for layout analysis, table recognition, and OCR.
- MinerU [10], an OCR-based tool from OpenDataLab that supports 84 languages.
- Marker [11], which combines layout detection with deep learning models.
- DeepSeek OCR [12], a Vision–Language Model approach.

These tools have been compared in various benchmarks. For example, OmniDocBench [9] evaluates parsing accuracy on 1355 pages using edit distance and table structure metrics. The proper Docling technical report [7] benchmarks conversion speed. Other tools and procedures have targeted French documents [13] or proposed unified evaluation frameworks [14].

Surprisingly, this set of benchmark measures does not assess whether better parsing actually leads to better RAG answers, which, a priori, would be a primary motivation for the benchmarking. In practical terms, a conversion tool might score well on text-fidelity metrics, which would be highly appreciated by end-users. Yet, it could produce a Markdown file that, once chunked and embedded, may fail to retrieve the correct passages for a given question. The reverse situation may also be a possibility. For example, a tool with messier output might still preserve enough semantic content to support accurate answers. This disconnect between parsing quality and downstream task performance is a gap in the current literature, which we address in the present study. To our knowledge, there is no prior work that systematically evaluates PDF conversion frameworks through the lens of RAG question-answering accuracy, which is the main aim of this study.

This article reports on the effort to explore that gap. We started by building a modular pipeline that supports four conversion frameworks (Docling, MinerU, Marker, and DeepSeek OCR). Since Marker's cloud version raises data privacy concerns, an obvious constraint when handling Portuguese Army documents, and its local version produced inferior results, Marker was excluded from the quantitative benchmark. Thus, we tested 19 configurations across the remaining frameworks, containing variations not only in the framework but also of the text cleaning steps, the document hierarchy rebuilding method, the chunking strategy, and the use of metadata enrichment (the addition of structural context, such as section paths and image descriptions, to each chunk). Finally, we evaluated all configurations against a benchmark of 50 manually crafted questions over a voluminous corpus of 36 PDF documents (1706 pages, roughly 492,000 words) from the military human resources administration written in Portuguese. Two baselines served as references for the results: a lower one using LangChain's PDFLoader with no preprocessing and an upper one using manually corrected Markdown files by the authors.

Beyond the extracted text, metadata plays a critical role in assisting the LLM during Chain-of-Thought reasoning, enabling it to filter irrelevant documents and sections before generating answers [15,16]. Thus, the pipeline's data repository was designed not only to store converted text but also to manage structured metadata alongside it.

The Medallion Architecture [17], with Bronze (raw PDFs), Silver (intermediate processing artifacts), and Gold (cleaned Markdown with metadata, ready for indexing) layers, inspired the pipeline's data repository's layered design. The big advantage is that this structure makes experiments reproducible. Switching from one framework to another requires changing a single configuration parameter that separates concerns among ingestion, transformation, and indexing.

In this study, the addition of a knowledge graph (GraphRAG) on top of the basic RAG pipeline was also explored. The expectation, supported by the recent literature on knowledge graph-enhanced RAG [6,18,19], was that introducing structured representations encoding the relationships and characteristics of the corpus would improve results. A graph with over 20,000 entities and 26,000 relationships was constructed using LangChain's LLMGraphTransformer to extract entity–relationship triples and store them in Neo4j. The outcome fell far short of expectations. With GraphRAG, the score on our benchmark was 82%, well below the 94.1% achieved by basic RAG with a well-produced corpus. In addition, entity deduplication via semantic similarity made things marginally worse (81%); i.e., it was not better. These negative findings are reported here because they may be instructive. The negative unexpected results suggest that naïve knowledge graph construction from LLM-extracted triples, without careful ontology design and graph densification, does not yet compete with straightforward vector-based retrieval when the underlying text is high-quality, which is also a major finding in the present study.

Prior work supports the intuition that preprocessing matters. Akbar et al. [20] demonstrated that a RAG system built on a curated legal corpus outperformed leading commercial LLMs across multiple evaluation metrics, underscoring the value of domain-specific document preparation. Yun and Lee [21] similarly showed that the careful filtering and structuring of legislative texts were essential to reliable LLM-based classification of transportation policy proposals. These studies, while focused on different domains, corroborate the general principle that the quality of the data pipeline preceding the LLM determines the quality of the output.

This paper addresses three research questions:

- RQ1: To what extent do PDF-to-Markdown conversion frameworks differ in the accuracy of downstream RAG question answering, when all other pipeline components are held constant?
- RQ2: Which pipeline dimensions, i.e., conversion framework, cleaning transformations, splitting strategy, or metadata enrichment, contribute the most to accuracy?
- RQ3: Can an automated preprocessing pipeline match or exceed the accuracy of manually curated document conversion?

Nevertheless, the main finding can be stated very simply: data preparation quality dominates and is of paramount importance. The accuracy gap between the worst con-figuration (79.1% with DeepSeek OCR) and the best automated one (94.1% with Docling plus hierarchical splitting and image descriptions) is nearly 15 percentage points. The gap between the naïve PDFLoader baseline (86.2%) and hand-corrected Markdowns (91.3%) exceeds 5 percentage points. In conclusion, enriching the metadata and, especially, applying hierarchy-aware chunking turned out to matter more than which conversion tool or framework was used. These conclusions draw special attention to the procedure of converting PDFs to text: overlooking the transformation of raw data into perfectly LLM-perceptible text may be at the origin of undesirable results in the LLM's answers, following the common adage: garbage in, garbage out (and the LLM does not have to be blamed).

The remainder of this article is organized as follows: Section 2, Methods and Materials, describes the system architecture, the document corpus, the evaluated frameworks, and the experimental methodology. Section 3 presents the benchmark results and the knowledge graph exploration. In Section 4, the implications and limitations of the findings are discussed. Finally, Section 5 concludes this study.

## 2. Materials and Methods

In this section, the system architecture, the document corpus, the evaluated PDF conversion frameworks, the evaluation dataset, and the experimental method are described.

### *2.1. System Architecture*

The system architecture is composed of layered data repositories inspired by the Medallion Architecture [17], which was adapted for RAG document management. The architecture separates raw document ingestion from transformation and indexing, enabling reproducible experimentation and an interchangeable framework. The layers in question are as follows:

- Bronze Layer (Landing Zone): It contains raw PDF documents and additional metadata in JSON files for semantic enrichment and file cataloging; in this layer, no transformations are applied.
- Silver Layer (Processing): It serves as an internal cleaning section where data are curated and enriched for final consumption; this layer does not have a predefined structure, allowing for flexibility in pipeline implementation.
- Gold Layer (RAG-Ready): It provides RAG-level data, where documents are organized into bundles that expand into a dedicated directory containing processed text as Markdown, extracted image assets, and metadata enrichment as JSON files.

This architecture does not include the knowledge graph, data indexing, or embedding vectors, which are considered an additional layer outside the proposed repository structure. A complete specification of the data repository is available in the project repository [22].

The system is made up of two main pipelines, ETL (Extract, Transform, Load) and indexing, an ingestion controller, and a query API (Application Programming Interface).

#### 2.1.1. ETL Pipeline

The ETL (Extract, Transform, Load) pipeline takes advantage of the system architecture, transforming PDFs from the Bronze layer into RAG-ready Markdown in the Gold layer. Before the ETL process begins, an organizing step hash-names the incoming PDFs, copies them into the Landing Zone (Bronze layer), and registers document information in the catalog. While the pipeline follows the ETL pattern, the stages are grouped as shown in Figure 1 and detailed below:

1. Extract: Convert PDF to Markdown using the configured converter, producing intermediate output in the Silver layer.
2. Transform: Apply configurable cleaning operations and hierarchy rebuilding.

3. Load: Save cleaned Markdown and extracted assets to the Gold layer.

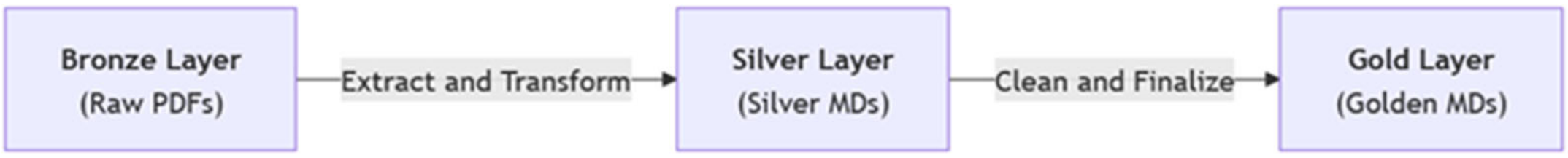

**Figure 1.** ETL pipeline workflow. Raw PDFs from the Bronze layer are extracted and transformed into intermediate Markdown (Silver layer) and then cleaned and finalized into RAG-ready Markdown with extracted assets (Gold layer).

The pipeline supports several configurable transformation options designed to address known issues in framework outputs: HTML table cleaning (converting HTML tables to Markdown tables), LaTeX formula cleaning (converting LaTeX equations to plain text), and hierarchy rebuilding (reconstructing document header levels using either font-based PDF analysis or LLM-based inference). The conversion framework itself is selectable through a single configuration parameter, enabling transparent switching between Docling, MinerU, Marker, or DeepSeek OCR (Optical Character Recognition) without code changes.

#### 2.1.2. Indexing Pipeline

The indexing pipeline processes documents from the Gold layer (or directly from the Bronze layer if the converter used is PDFLoader, bypassing the ETL) into a vector store (a database that stores embedded text segments and supports similarity-based retrieval) through four sequential stages:

1. Load: Read Markdown from the Gold layer or PDFs from the Bronze layer.
2. Split: Chunk documents using the configured splitting strategy (the method used to divide documents into chunks for indexing).
3. Embed: Generate embeddings via OpenAI's text-embedding-3-small model.
4. Store: Upsert into a ChromaDB vector store.

Three splitting strategies were created to support the framework:

- Recursive: Basic character-based splitting with overlap, using a sliding window approach without awareness of document structure; this strategy is the fastest, although it does not attend to semantic boundaries and, therefore, is not recommended, attending to the issues raised in the Introduction.
- Markdown Recursive: Section-aware chunking that respects Markdown structure (headers, lists, and code blocks) when creating chunks, providing better semantic cohesion than recursive splitting.
- Hierarchical Recursive: It leverages the full document hierarchy (headers, sections, and subsections) to create semantically meaningful chunks. Unlike the other two strategies, chunks are bounded by section markers (Markdown headers), so a chunk that falls between two hierarchy boundaries may be shorter than the configured maximum size. Each chunk is prepended with breadcrumb context (the path through the document tree from root to the current section); this strategy requires well-structured Markdown with clear hierarchy levels and is currently only fully supported by Docling and hand-made outputs.

#### 2.1.3. Ingestion Orchestrator

Through a settings file, this system's component orchestrates the entire process of invoking the required frameworks, transformation mechanisms, and sequence workflows. This component is of major importance, as it enables fast and easy workflow reconfiguration. For instance, it enables switching from using the LangChain PDFLoader (LangChain ecosystem version 1.0.0; langchain-community version 0.4; pypdf version 6.1.1) to a workflow applying a specific framework and RAG indexing.

#### 2.1.4. System API

A query handler provides a REST (Representational State Transfer) API interface that enables third-party applications to submit natural language queries to the knowledge base and receive meaningful responses with source document references.

The complete system is containerized as a Docker image for user-friendly deployment. The system also includes a web-based chatbot interface that uses this API for user interaction, as well as a benchmarking tool for evaluating system performance. The chatbot allows users to submit queries in natural language and receive responses grounded in the document corpus, with references to the source documents from which the answer was derived.

### *2.2. Document Corpus*

The corpus consists of 36 publicly accessible PDF documents provided by the human resources management service of the Personnel Command of the Portuguese Army, totaling 1706 pages and 491,562 words (3,154,199 characters). The corpus size is 84.6 MB, and all documents are written in European Portuguese. Table 1 summarizes the corpus descriptive statistics.

**Table 1.** Corpus overview and descriptive statistics.

| Metric | Value |
|---|---|
| Total documents | 36 |
| Total pages | 1706 |
| Total words | 491,562 |
| Total characters | 3,154,199 |
| Total embedded images | 4198 |
| Unique images (after deduplication) | 1123 |
| Image duplication rate | 73.2% |
| Corpus size | 84.6 MB |
| Language | Portuguese (100%) |
| Documents with form fields | 16 (44.4%) |
| Documents with tabular content | 34 (94.4%) |
| Pages per document (mean ± std) | 47.39 ± 53.21 |
| Pages per document (median) | 32 |
| Pages per document (range) | 1–244 |
| Words per document (mean ± std) | 13,654.5 ± 23,938.1 |
| Words per document (median) | 5886 |
| Vocabulary richness (mean) [1] | 0.21 |
| Average word length (characters) | 5.73 |

[1] Vocabulary richness is the ratio of unique words to total words per document.

The documents span eight categories reflecting diverse structural characteristics: legal consolidations (7 documents, 725 pages), manuals (6 documents, 485 pages), procedures (2 documents, 121 pages), brochures (3 documents, 108 pages), information notes (6 documents, 96 pages), guides (5 documents, 62 pages), legal decrees (5 documents, 55 pages), and other (2 documents, 54 pages). Table 2 presents the distribution by category.

Several characteristics of the corpus make it challenging for automated conversion. Nearly all documents (94.4%) contain tabular content, often with merged rows and columns. Close to half (44.4%) contain interactive form fields. The documents contain 4198 images, of which only 1123 (26.8%) are unique after deduplication. These include photographs (37.2%), diagrams and charts (4.7%), screenshots (9.0%), logos (14.9%), and decorative elements (14.6%). Furthermore, the documents were produced by at least eight different PDF tools (including Microsoft Word 2016, Acrobat Distiller, iText, Ghostscript, and others), reflecting heterogeneous origins that further complicate consistent parsing.

**Table 2.** Document distribution by category.

| Category | Docs | Pages | Words | Size (MB) | Words/Page |
|---|---|---|---|---|---|
| Legal consolidations | 7 | 725 | 332,376 | 13.9 | 458.4 |
| Manuals | 6 | 485 | 50,602 | 33.6 | 104.3 |
| Procedures | 2 | 121 | 29,964 | 1.3 | 247.6 |
| Brochures | 3 | 108 | 9363 | 13.8 | 86.7 |
| Information notes | 6 | 96 | 24,714 | 7.1 | 257.4 |
| Guides | 5 | 62 | 7490 | 5.8 | 120.8 |
| Legal decrees | 5 | 55 | 29,265 | 3.7 | 532.1 |
| Other | 2 | 54 | 7788 | 5.4 | 144.2 |
| Total | 36 | 1706 | 491,562 | 84.6 | 288.1 |

The text density varies considerably across categories: legal decrees and legal consolidations are text-dense (458 to 532 words per page), while brochures and manuals are image-heavy (87 to 104 words per page). This variation tests the frameworks' ability to handle both text-heavy and visually rich documents.

### *2.3. Evaluated Frameworks*

Four open-source PDF conversion frameworks were evaluated, selected to represent the diversity of approaches currently available: pipeline-based tools using specialized models, OCR-based tools, VLM-based tools, and cloud-based services. Table 3 summarizes their characteristics, strengths, and limitations as observed in our experiments.

**Table 3.** Evaluated converter frameworks: strengths and limitations observed.

| Framework | Approach | Strengths | Limitations | Best Accuracy |
|---|---|---|---|---|
| PDFLoader [23] | LangChain loader, direct text extraction | Simple, fast integration; useful as baseline | Loses document structure; tables incorrectly loaded with merged cells; low metadata control | 86.2% (config. 1, Recursive, K = 200) |
| MinerU Pipeline [10] | Local OCR-based conversion | Converts to Markdown/JSON; preserves headings and lists; extracts images and tables; handles scanned PDFs | Document hierarchy lost; some tables have wrong structure | 84.0% (config. 7, HR-LLM, K = 200) |
| MinerU HTTP-client VLM [10] | Web service with VLM | Same strengths with better table results than the Pipeline version | High hardware requirements; Portuguese "ç" misinterpreted; document hierarchy lost; system crashes on certain PDFs | 83.2% (config. 11, full cleaning, K = 200) |
| Marker (Cloud) [11] | Cloud-based conversion | Excellent results; complex tables well handled; hierarchy preserved | Cloud-only (privacy concerns); high computational requirements; local version produces inferior results | Not evaluated in benchmark (cloud-only constraint) |
| DeepSeek OCR [12] | VLM-based OCR | Great precision extracting native text and complex tables; successfully handled all documents | Requires page-to-image conversion; loses document structure; high hardware requirements | 79.1% (config. 21, HTML, K = 200) |
| Docling [7,8] | Modular pipeline with specialized models | Excellent Markdown conversion with Docling-hierarchical-pdf; complex tables without errors; hierarchy preserved; LangChain integration; runs locally; supports PDF, DOCX, HTML, PPTX; supports image analysis for descriptive text through an external VLM | Sometimes mistakes the hierarchy of headings, which requires additional visual checking | 94.1% (config 20, Hierarchical + Images, K = 200) |

In addition to the six conversion variations listed, we implemented all combinations in a configuration-driven pipeline, allowing for transparent selection through the ingestion orchestrator previously discussed. This design was instrumental for the systematic benchmarking described in Section 3, Results.

### 2.4. Evaluation Dataset

A benchmark dataset of 50 questions with expected answers was manually constructed to evaluate the downstream RAG accuracy. The questions were designed to test specific scenarios relevant to the corpus characteristics:

- Questions requiring data extraction from tables, including tables with merged rows or columns.
- Questions targeting text segments known to be incorrectly read by LangChain's PDFLoader.
- Questions requiring information from specific document sections, testing whether document hierarchy is preserved.
- Questions whose answers depend on the content of images or diagrams.

The expected answers were determined by the research team through the analysis of the source documents, ensuring a ground truth answer independent of any automated processing.

### 2.5. Experimental Configuration

All experiments shared a fixed RAG configuration to isolate the effect of the data preparation pipeline from other variables:

- LLM: gpt-4o-mini (OpenAI, San Francisco, CA, USA).
- Embedding model: text-embedding-3-small (OpenAI).
- Chunk size: 1000 characters (maximum), with chunk overlap of 200 characters; for the hierarchical recursive strategy, chunks may be shorter than 1000 characters when bounded by section markers in the document hierarchy.
- Default retrieval K: 200 top-k chunks.

The high value of K (200) was deliberately chosen to ensure that relevant information is available in the retrieval context for most queries. This design decision prioritizes the evaluation of pipeline effectiveness over retrieval precision. At K = 200, the retriever is unlikely to miss relevant chunks, so accuracy differences predominantly reflect the quality of the converted and indexed content rather than retrieval limitations. For this reason, we did not apply end-to-end RAG evaluation frameworks such as RAGAs [24] at this stage.

### 2.6. Evaluation Methodology

The accuracy of each pipeline configuration was evaluated using an LLM-as-judge (in which a separate LLM evaluates response quality) approach [25]. A prompt was provided to the LLM judge specifying the evaluation criteria to be applied when comparing the RAG response against the expected answer for each of the 50 questions. Each question received a score on a qualiquanti continuous scale from 0 to 100 distributed across five categories (cf. Appendix A).

To quantify the stability of these scores and enable statistical comparison across configurations, each configuration was evaluated over 50 independent runs of the full 50-question benchmark. In each run, the same set of questions was submitted to the pipeline and scored by the LLM judge, yielding a run mean (the average of the 50 question scores in that run). The reported accuracy for each configuration is the grand mean across these 50 run means, accompanied by the between-run standard deviation (SD) and a 95% confidence interval (CI). The between-run SD reflects the stochastic variability of the evaluation process across repeated runs. It should not be confused with within-run variability across questions, which reflects the heterogeneity of question difficulty and is substantially larger.

Confidence intervals were computed using the t-distribution with 49 degrees of freedom: CI = grand mean $\pm$ t49, 0.025 $\times$ SEM (standard error of the mean), where SEM = SD/$\sqrt{50}$. In practice, the resulting intervals are narrow (typically under 1 percentage point in width), indicating that LLM-as-judge evaluation is highly stable across runs.

Since the same 50 questions are used across all configurations, the run means are naturally paired by run index, making a paired design appropriate for statistical comparison. The normality of paired differences was assessed using the Shapiro–Wilk test (Appendix B). Since the majority of paired difference distributions depart from normality (18 of 24 comparisons), the Wilcoxon signed-rank test was adopted as the primary test of statistical significance. Effect sizes are reported using Cohen's d, computed as the mean of the paired differences divided by their standard deviation. Although Cohen's d assumes normality, it is used here as a descriptive measure of effect magnitude, which is standard practice when paired with a non-parametric significance test.

### 2.7. Pipeline Configurations

A total of 21 pipeline configurations were evaluated, systematically varying the following dimensions:

- Data preparation framework: PDFLoader (lower baseline), MinerU Pipeline, MinerU HTTP-client VLM, Hand-made Markdowns (upper baseline), Docling and DeepSeek OCR.
- Transformations: None, HTML table cleaning, LaTeX formula cleaning, font-based hierarchy rebuilding (HR-F), LLM-based hierarchy rebuilding (HR-LLM), VLM-based image descriptions, and combinations thereof.
- Splitting strategy: Recursive, Markdown recursive, or hierarchical recursive.
- Hierarchical metadata: Whether breadcrumb metadata about document structure is added to chunks (Yes/No).
- Retriever K: 200 (default), 50, 20, or 5 (tested with hand-made Markdowns and Docling to assess the interaction between pipeline quality and retrieval depth, i.e., the number of chunks retrieved and provided to the LLM as context); the default value of K = 200 was selected to ensure that retrieval failures did not confound the evaluation of pipeline quality.

Two baselines frame the evaluation: a lower baseline using LangChain's PDFLoader with no preprocessing (representing the minimal-effort approach) and an upper baseline using hand-made Markdowns, where all 36 documents were manually converted to Markdown (representing the best achievable result for the given RAG configuration). These baselines bound the interval within which automated frameworks are expected to perform.

### 2.8. Knowledge Graph Construction

In addition to the basic RAG pipeline, we conducted an exploratory study on knowledge graph construction for GraphRAG. Models were run locally on an NVIDIA DGX Spark workstation using LM Studio (version v0.0.5+2). The embedding model was BAAI/bge-m3 (identified as "text-embedding-bge-m3" in LM Studio) and the LLM was openai/gpt-oss-120b [26].

Markdown documents were divided into chunks using a sliding window technique (1000 characters, 200-character overlap). Each chunk was stored in a Neo4j graph database as a TextChunk node, containing the original text, embedding vector, and source metadata (document ID and position).

Entity extraction was performed using LangChain's LLMGraphTransformer class, which analyses text chunks and extracts triples (entity–relationship–entity). To avoid entity duplication, Neo4j's MERGE logic was used. If an entity already exists, the system creates new relationships or updates properties rather than duplicating the node.

Data are stored using a double-layer approach, as illustrated in Figure 2. TextChunk nodes store the chunk identifier, text content, source document reference (document ID and path), chunk position index, and embedding vector. Entity nodes store a unique identifier, the original entity name, and a semantic type (e.g., Person, Organization, Location, Document, Process, and Role). Two relationship types connect the nodes: MENTIONS links each TextChunk to the entities extracted from it (one-to-many), while RELATED captures semantic relationships between entities themselves.

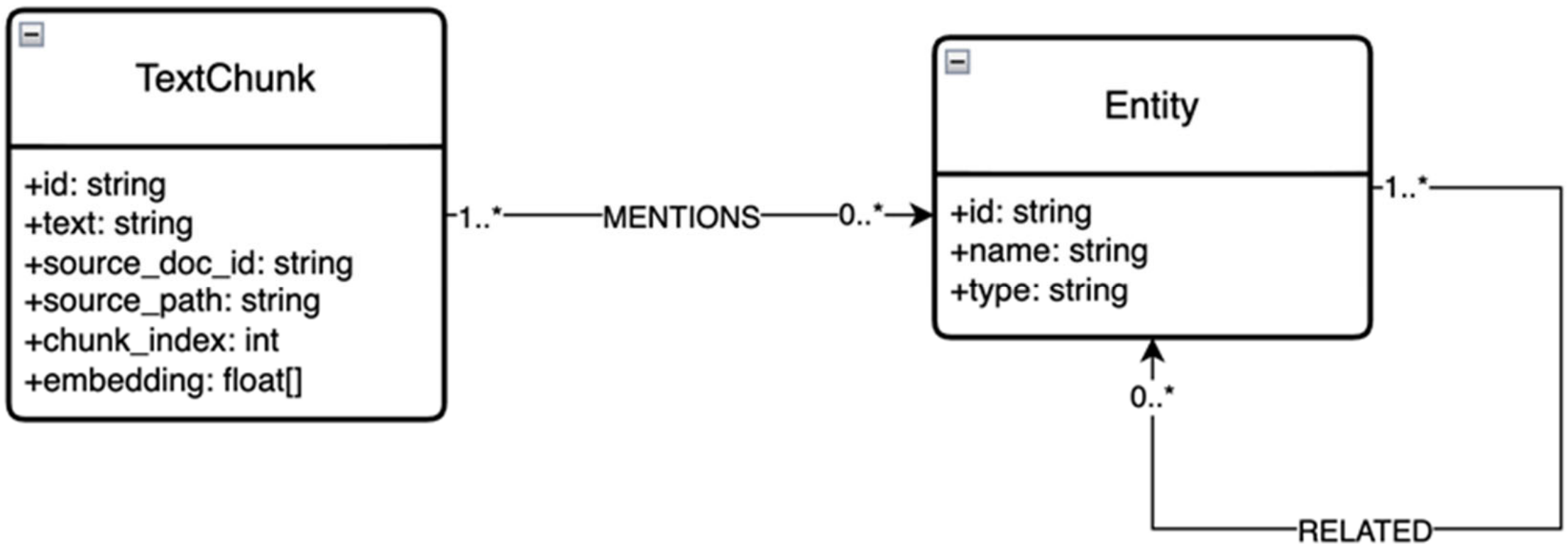

**Figure 2.** Knowledge graph data model. TextChunk nodes store text content with source metadata and embedding vectors. Entity nodes store a unique identifier, name, and semantic type. MENTIONS relationships link text chunks to the entities extracted from them. RELATED relationships capture semantic connections between entities. * means multiple.

A semantic deduplication pipeline was subsequently applied to address entity duplication arising from linguistic variations (e.g., singular/plural forms and synonyms). This pipeline identifies and merges semantically similar entities like the MERGE logic using embeddings and cosine similarity, with entities exceeding 85% similarity merged into canonical forms.

## 3. Results

This section presents the results of the experiments organized in three parts. Firstly, we begin with the benchmark results across the 19 pipeline configurations (Section 3.1), followed by a more detailed analysis of the factors that influenced accuracy (Section 3.2), and conclude with the results of the knowledge graph exploration (Section 3.3).

### *3.1. Pipeline Benchmark*

Table 4 presents the benchmark results for all 21 configurations tested. The table includes two baselines: configuration 1 (PDFLoader, the lower baseline) and configurations 12 to 15 (hand-made Markdowns at different retrieval depths, serving as the upper baseline). The remaining configurations represent automated pipelines using MinerU, Docling, and DeepSeek OCR with various combinations of transformations and splitting strategies. Marker was not tested since the cloud version lacks privacy, an obvious requirement from the Portuguese Army, and the local option was underachieving.

The results achieved with the PDFLoader baseline (86.2 ± 0.9%) were higher than initially expected. The hand-made Markdown baseline with K = 200 reached 91.3 ± 1.2%, and with K = 50, it reached 92.9 ± 0.4%. It should be noted that the hand-made baseline, while representing the best result achievable through manual curation for the given RAG configuration, is not a theoretical ceiling. It reflects the quality of manual Markdown conversion, which is itself imperfect.

**Table 4.** Benchmark results across 21 pipeline configurations. Accuracy is reported as the grand mean ± between-run SD [95% CI] over 50 independent evaluation runs. K denotes the number of retrieved chunks. Configurations are grouped by data preparation framework.

| # | Data Preparation | K | Transformations | Splitting | Hier. Meta | Accuracy (%) Mean ± SD [95% CI] |
|---|---|---|---|---|---|---|
| 1 | PDFLoader | 200 | — | Recursive | No | 86.2 ± 0.9 [86.0, 86.5] |
| 2 | MinerU Pipeline | 200 | — | Recursive | No | 82.8 ± 1.3 [82.4, 83.2] |
| 3 | MinerU Pipeline | 200 | — | Markdown Recurs. | No | 83.5 ± 1.4 [83.1, 83.9] |
| 4 | MinerU Pipeline | 200 | HTML | Markdown Recurs. | No | 83.0 ± 0.7 [82.8, 83.2] |
| 5 | MinerU Pipeline | 200 | LaTeX | Markdown Recurs. | No | 83.9 ± 0.5 [83.8, 84.1] |
| 6 | MinerU Pipeline | 200 | HR-F | Markdown Recurs. | No | 83.0 ± 0.9 [82.7, 83.2] |
| 7 | MinerU Pipeline | 200 | HR-LLM | Markdown Recurs. | No | 84.0 ± 0.7 [83.8, 84.1] |
| 8 | MinerU Pipeline | 200 | HTML, LaTeX, HR-F | Markdown Recurs. | No | 81.5 ± 0.7 [81.4, 81.7] |
| 9 | MinerU Pipeline | 200 | HTML, LaTeX, HR-LLM | Markdown Recurs. | No | 83.9 ± 1.0 [83.6, 84.2] |
| 10 | MinerU HTTP-client VLM | 200 | — | Markdown Recurs. | No | 82.1 ± 0.5 [82.0, 82.3] |
| 11 | MinerU HTTP-client VLM | 200 | HTML, LaTeX, HR-F | Markdown Recurs. | No | 83.2 ± 0.8 [83.0, 83.4] |
| 12 | Hand-made Markdowns | 200 | — | Hierarchical Recurs. | Yes | 91.3 ± 1.2 [91.0, 91.7] |
| 13 | Hand-made Markdowns | 50 | — | Hierarchical Recurs. | Yes | 92.9 ± 0.4 [92.8, 93.0] |
| 14 | Hand-made Markdowns | 20 | — | Hierarchical Recurs. | Yes | 87.6 ± 1.7 [87.1, 88.0] |
| 15 | Hand-made Markdowns | 5 | — | Hierarchical Recurs. | Yes | 78.4 ± 0.6 [78.2, 78.5] |
| 16 | Docling | 200 | — | Recursive | no | 84.7 ± 1.8 [84.2, 85.2] |
| 17 | Docling | 50 | — | Recursive | No | 82.9 ± 1.2 [82.6, 83.3] |
| 18 | Docling | 5 | — | Recursive | No | 82.5 ± 0.9 [82.2, 82.7] |
| 19 | Docling | 200 | — | Hierarchical Recurs. | Yes | 91.2 ± 1.0 [90.9, 91.5] |
| 20 | Docling | 200 | Images | Hierarchical Recurs. | Yes | 94.1 ± 1.6 [93.6, 94.5] |
| 21 | DeepSeek OCR | 200 | HTML | Markdown Recurs. | No | 79.1 ± 1.1 [78.8, 79.4] |

Among the automated frameworks, Docling with hierarchical recursive splitting and VLM-based image descriptions (configuration 20) achieved the highest accuracy at 94.1 ± 1.6%, surpassing the hand-made upper baseline. This result indicates that Docling's hierarchical structure extraction, combined with metadata enrichment, can match and even exceed the quality of manual curation for this corpus. At the other extreme, DeepSeek OCR (configuration 21) scored 79.1 ± 1.1%, the lowest result in the benchmark when comparing configurations at K = 200. The span between the worst and best automated configurations at K = 200 is approximately 15 percentage points, confirming that data preparation choices dominate RAG accuracy in our experimental setting.

The large majority of pairwise differences between configurations are statistically significant (Wilcoxon signed-rank tests, $p < 0.05$; cf. Table 5). Effect sizes range from small ($|d| \approx 0.3$ for individual MinerU transformation variants in Section 3.2.2) to very large ($d > 4.0$ for the hierarchical splitting comparison in Section 3.2.3). The single comparison that does not reach significance is Docling at K = 50 vs. K = 5 ($p = 0.090$, d = 0.28), consistent with the observation that retrieval depth has diminishing returns without hierarchical metadata (Section 3.2.4). The confidence intervals in Table 4, none of which exceed 1 percentage point in width, confirm that the observed differences reflect genuine performance gaps rather than stochastic variation in the LLM-as-judge evaluation. Appendix B reports the complete list of the Shapiro–Wilk normality tests.

### 3.2. Analysis of Contributing Factors

In order to understand what drives these accuracy differences, we examined the results from several angles: the effect of the conversion framework itself, the role of cleaning transformations, the contribution of splitting strategy and metadata enrichment, and the interaction between pipeline quality and retrieval depth.

**Table 5.** Pairwise statistical comparisons for key configuration pairs. All tests use 50 paired runs. Configurations = configuration numbers. Δ = mean difference in percentage points. $p$ = $p$-value from the Wilcoxon signed-rank test. Cohen's d effect size. † Not significant at $\alpha = 0.05$.

| Section | Comparison | Configurations | | Δ (pp) | $p$ (Wilcoxon) | Cohen's d |
|---|---|---|---|---|---|---|
| 3.2.1 | PDFLoader vs. MinerU (Recursive) | 1 | 2 | +3.4 | $7.5 \times 10^{-10}$ | +1.92 |
| 3.2.1 | PDFLoader vs. Docling (Recursive) | 1 | 16 | +1.5 | $4.6 \times 10^{-7}$ | +0.71 |
| 3.2.1 | Docling vs. MinerU (Recursive) | 16 | 2 | +1.9 | $2.8 \times 10^{-9}$ | +1.58 |
| 3.2.1 | MinerU: Recursive vs. Md Recursive | 2 | 3 | −0.7 | $1.4 \times 10^{-4}$ | −0.68 |
| 3.2.2 | MinerU: base vs. HTML | 3 | 4 | +0.5 | $1.1 \times 10^{-4}$ | +0.31 |
| 3.2.2 | MinerU: base vs. LaTeX | 3 | 5 | −0.5 | 0.014 | −0.31 |
| 3.2.2 | MinerU: base vs. HR-F | 3 | 6 | +0.5 | 0.006 | +0.27 |
| 3.2.2 | MinerU: base vs. HR-LLM | 3 | 7 | −0.5 | 0.014 | −0.32 |
| 3.2.2 | MinerU: HR-F vs. HR-LLM (indiv.) | 6 | 7 | −1.0 | $6.5 \times 10^{-6}$ | −0.78 |
| 3.2.2 | MinerU: HR-F vs. HR-LLM (comb.) | 8 | 9 | −2.4 | $1.1 \times 10^{-9}$ | −1.73 |
| 3.2.3 | Docling: Recursive vs. Hierarchical | 16 | 19 | −6.5 | $7.6 \times 10^{-10}$ | −4.54 |
| 3.2.3 | Docling: Hierarchical vs. +Images | 19 | 20 | −2.9 | $2.0 \times 10^{-9}$ | −1.86 |
| 3.2.3 | Docling: Recursive vs. Hier. + Images | 16 | 20 | −9.4 | $7.6 \times 10^{-10}$ | −5.69 |
| 3.2.3 | Docling + Img vs. Hand-made (K = 200) | 20 | 12 | +2.7 | $8.2 \times 10^{-8}$ | +1.18 |
| 3.2.3 | Docling + Img vs. Hand-made (K = 50) | 20 | 13 | +1.2 | $5.9 \times 10^{-7}$ | +0.71 |
| 3.2.4 | Hand-made: K = 50 vs. K = 200 | 13 | 12 | +1.5 | $7.0 \times 10^{-9}$ | +1.18 |
| 3.2.4 | Hand-made: K = 50 vs. K = 20 | 13 | 14 | +5.3 | $7.5 \times 10^{-10}$ | +2.98 |
| 3.2.4 | Hand-made: K = 20 vs. K = 5 | 14 | 15 | +9.2 | $7.6 \times 10^{-10}$ | +5.12 |
| 3.2.4 | Docling: K = 200 vs. K = 50 | 16 | 17 | +1.8 | $8.3 \times 10^{-7}$ | +0.74 |
| 3.2.4 | Docling: K = 50 vs. K = 5 | 17 | 18 | +0.5 | 0.090 † | +0.28 |
| 3.2.5 | MinerU Pipeline vs. VLM (no transf.) | 3 | 10 | +1.3 | $1.4 \times 10^{-8}$ | +0.94 |
| 3.2.5 | MinerU Pipeline vs. VLM (full clean.) | 8 | 11 | −1.6 | $2.0 \times 10^{-9}$ | −1.45 |
| Overall | Best vs. Worst (Docling vs. DeepSeek) | 20 | 21 | +15.0 | $1.8 \times 10^{-15}$ | +7.68 |
| Overall | PDFLoader vs. Docling + Images | 1 | 20 | −7.8 | $8.0 \times 10^{-10}$ | −3.92 |

#### 3.2.1. Effect of the Conversion Framework

Comparing the three automated frameworks using recursive splitting at K = 200, with no additional transformations or metadata, PDFLoader (config. 1: 86.2 ± 0.9%) achieved the highest accuracy, followed by Docling (config. 16: 84.7 ± 1.8%) and MinerU Pipeline (config. 2: 82.8 ± 1.3%). All three pairwise differences are statistically significant (Wilcoxon signed-rank test, $p < 10^{-6}$; Cohen's d = 0.71 to 1.92).

Switching MinerU from recursive to Markdown recursive splitting (config. 2 → 3) yields a modest improvement from 82.8% to 83.5% ($p < 10^{-4}$, d = 0.68). The individual cleaning transformations tested on MinerU output (Section 3.2.2) produce similarly small effects at K = 200. Larger gains are observed when Docling is combined with hierarchical splitting (Section 3.2.3).

#### 3.2.2. Impact of Cleaning Transformations

We tested several transformations on MinerU Pipeline output, all with Markdown recursive splitting at K = 200, whose results are as follows:

- The baseline for comparison, i.e., with no transformations (config. 3), reached 83.5 ± 1.4%.
- HTML table cleaning (config. 4) is similar to the baseline comparison, with 83.0 ± 0.7%.
- LaTeX cleaning (config. 5) is also similar, with an accuracy of 83.9 ± 0.5%.
- Font-based hierarchy rebuilding, HR-F (config. 6), repeats the pattern, with 83.0 ± 0.9%.
- LLM-based hierarchy rebuilding, HR-LLM (config. 7), also repeats the pattern, at 84.0 ± 0.7%.

The individual transformations produce differences of less than 1 percentage point relative to the untransformed baseline (config. 3). Of these, the HR-F comparison, while statistically significant by the Wilcoxon signed-rank test ($p = 0.006$), shows a small effect size (d = 0.27).

When transformations are combined, the results diverge:

- HTML + LaTeX + HR-F (config. 8) scores 81.5 ± 0.7%, the lowest among all MinerU configurations.
- HTML + LaTeX + HR-LLM (config. 9) scores 83.9 ± 1.0%, on par with the individual transformations.

The difference between configs. 8 and 9 is statistically significant ($p < 10^{-16}$, d = 1.73). Comparing the two hierarchy rebuilding approaches directly, HR-LLM outperforms HR-F, both when applied individually (config. 7 vs. 6: +1.0 pp, $p < 10^{-6}$, d = 0.78) and when combined with other transformations (config. 9 vs. 8: +2.4 pp, $p < 10^{-16}$, d = 1.73).

#### 3.2.3. Contribution of Splitting Strategy and Metadata Enrichment

The comparison between Docling configurations reveals the impact of the splitting strategy and hierarchical metadata. Moving from recursive splitting without metadata (config. 16: 84.7 ± 1.8%) to hierarchical recursive splitting with breadcrumb metadata (config. 19: 91.2 ± 1.0%) yields a 6.5-percentage-point improvement ($p < 10^{-34}$, d = 4.54) without utilizing any transformation step. This is the largest single-factor gain observed in the benchmark.

Adding VLM-based image descriptions (config. 19 → 20) contributes with an additional 2.9 percentage points, reaching 94.1 ± 1.6% ($p < 10^{-17}$, d = 1.86). This configuration surpasses the hand-made Markdown upper baseline at K = 200 (config. 12: 91.3 ± 1.2%; difference: +2.8 pp, $p < 10^{-10}$, d = 1.18) and also exceeds the hand-made baseline at K = 50 (config. 13: 92.9 ± 0.4%; difference: +1.2 pp, $p < 10^{-5}$, d = 0.71).

#### 3.2.4. Interaction Between Pipeline Quality and Retrieval Depth

We tested multiple values of K (the number of retrieved chunks) with hand-made Markdowns (configs. 12 to 15) and with Docling using recursive splitting and no metadata (configs. 16 to 18). For hand-made Markdowns:

- K = 200 → 91.3 ± 1.2%.
- K = 50 → 92.9 ± 0.4%.
- K = 20 → 87.6 ± 1.7%.
- K = 5 → 78.4 ± 0.6%.

Accuracy increases sharply from K = 5 to K = 50 (+14.5 pp, $p < 10^{-37}$) but slightly decreases from K = 50 to K = 200 (−1.5 pp, $p < 10^{-10}$).

For Docling (recursive, no metadata):

- K = 200 → 84.7 ± 1.8.
- K = 50 → 82.9 ± 1.2%.
- K = 5 → 82.5 ± 0.9%.

The difference between K = 5 and K = 200 is 2.2 percentage points ($p < 10^{-5}$, d = 0.74). The difference between K = 50 and K = 5 is 0.4 pp and does not reach statistical significance (Wilcoxon $p = 0.090$).

#### 3.2.5. MinerU HTTP-Client VLM

The MinerU HTTP-client VLM variant, which adds a Vision–Language Model for processing, did not yield the improvements anticipated. Without transformations (config. 10: 82.1 ± 0.5%), it scored below the Pipeline version with the same splitting strategy (config.

3: 83.5 ± 1.4%; difference: −1.4 pp, $p < 10^{-8}$, d = 0.94). With full cleaning (config. 11: 83.2 ± 0.8%), it outperformed the equivalent Pipeline configuration (config. 8: 81.5 ± 0.7%; difference: +1.6 pp, $p < 10^{-13}$, d = 1.45).

We should also note that the VLM variant processed only 24 of the 36 documents in the corpus, since it crashed on the remaining 12, forcing us to fall back to the Pipeline version for those documents. This instability, combined with the higher computational requirements and the misrecognition of Portuguese-specific characters, makes the HTTP-client VLM variant impractical for our use case.

#### 3.2.6. Performance by Question Type

To examine how preprocessing choices affect different types of questions, we categorized the 50 benchmark questions into four groups based on the knowledge and retrieval strategy required: table-dependent questions (9 questions) that require extracting specific values from tabular data; hierarchy-dependent questions (6 questions) that require navigating the document section structure to locate the correct answer; system/SAP questions (12 questions) that target specific SAP transaction codes or infotypes described in technical manuals; and straightforward text questions (23 questions) that can be answered from running text without requiring table or hierarchy navigation. Table 6 presents the mean accuracy by question type across selected configurations.

**Table 6.** Mean accuracy (%) by question type across selected configurations. Each value is the average score across all questions of that type, averaged over 50 independent runs. N = number of questions in each category.

| Question Type | N | PDFLoader (Config 1) | MinerU (Config 2) | Hand-Made (Config 12) | Docling Recursive (Config 16) | Docling Hierarchic. (Config 19) | Docling + Images (Config 20) | DeepSeek (Config 21) |
|---|---|---|---|---|---|---|---|---|
| Table-dependent | 9 | 70.1 | 69.5 | 99.9 | 66.6 | 99.8 | 99.5 | 80.2 |
| Hierarchy-dependent | 6 | 99.7 | 93.3 | 87.9 | 90.3 | 88.8 | 92.6 | 52.2 |
| System/SAP | 12 | 85.4 | 78.3 | 83.5 | 87.7 | 88.9 | 94.7 | 88.2 |
| Straightforward text | 23 | 89.5 | 87.6 | 93.0 | 88.8 | 89.6 | 92.0 | 80.9 |
| Overall | 50 | 86.2 | 82.8 | 91.3 | 84.7 | 91.2 | 94.1 | 79.1 |

The most striking pattern is the performance on table-dependent questions. PDFLoader (70.1%), MinerU (69.5%), and Docling with recursive splitting (66.6%) all score equal to or below 70% on this category, while Docling with hierarchical splitting (99.8%), the hand-made baseline (99.9%), and Docling with images (99.5%) score near-perfect. The 33-percentage-point gain from recursive to hierarchical splitting within Docling on table-dependent questions (66.6% → 99.8%) far exceeds the corresponding gains on the other categories (0.8 to 1.3 pp). This indicates that the hierarchical splitting strategy is primarily responsible for correct table retrieval and that the overall accuracy gains reported in Section 3.2.3 are largely driven by this question type.

DeepSeek OCR scores notably low on hierarchy-dependent questions (52.2%), reflecting its inability to preserve document structure during conversion. On straightforward text questions, all frameworks score between 80.9% and 93.0%, with smaller differences between configurations. System/SAP questions show a more gradual progression across configurations, with Docling + images achieving the highest score (94.7%).

### 3.3. Knowledge Graph Exploration

Given the quality of the data in the best Gold layer, we proceeded to explore whether a knowledge graph could improve retrieval beyond what the basic RAG pipeline achieved.

We used LangChain's LLMGraphTransformer to extract entity–relationship triples from the text chunks and stored them in a Neo4j database.

#### 3.3.1. Initial Graph Construction

The initial graph construction over the full corpus produced 5899 TextChunk nodes, 20,055 entity nodes, and 26,067 relationships. The graph has the complexity of a large graph, but without the knowledge density that would make GraphRAG effective, particularly when applying community detection algorithms such as Leiden. For the purpose of confirming this assessment, we evaluated our GraphRAG implementation using the same 50-question benchmark, obtaining an accuracy of 82.0%. This is well below the 94.1% achieved by basic RAG with Docling and image descriptions (config. 20) and even below the PDFLoader baseline (86.2%).

#### 3.3.2. Entity Deduplication

One of the first limitations we identified in the initial graph concerned entity duplication arising from linguistic variations. For instance, the same concept might appear as both singular and plural forms or as synonyms with slightly different surface forms. To address this, we implemented a semantic deduplication pipeline that computes embeddings for all entities and merges those with cosine similarity above 85%.

This process reduced the entity count from 20,055 to 17,494 canonical entities, while the number of relationships decreased slightly to 25,674. Figure 3 depicts the resulting knowledge graph after deduplication, visualized using 3d-force-graph. The visualization reveals a dense central cluster with numerous peripheral nodes, organized into 1408 clusters. However, the deduplication did not improve GraphRAG accuracy. It actually even decreased marginally from 82% to 81%.

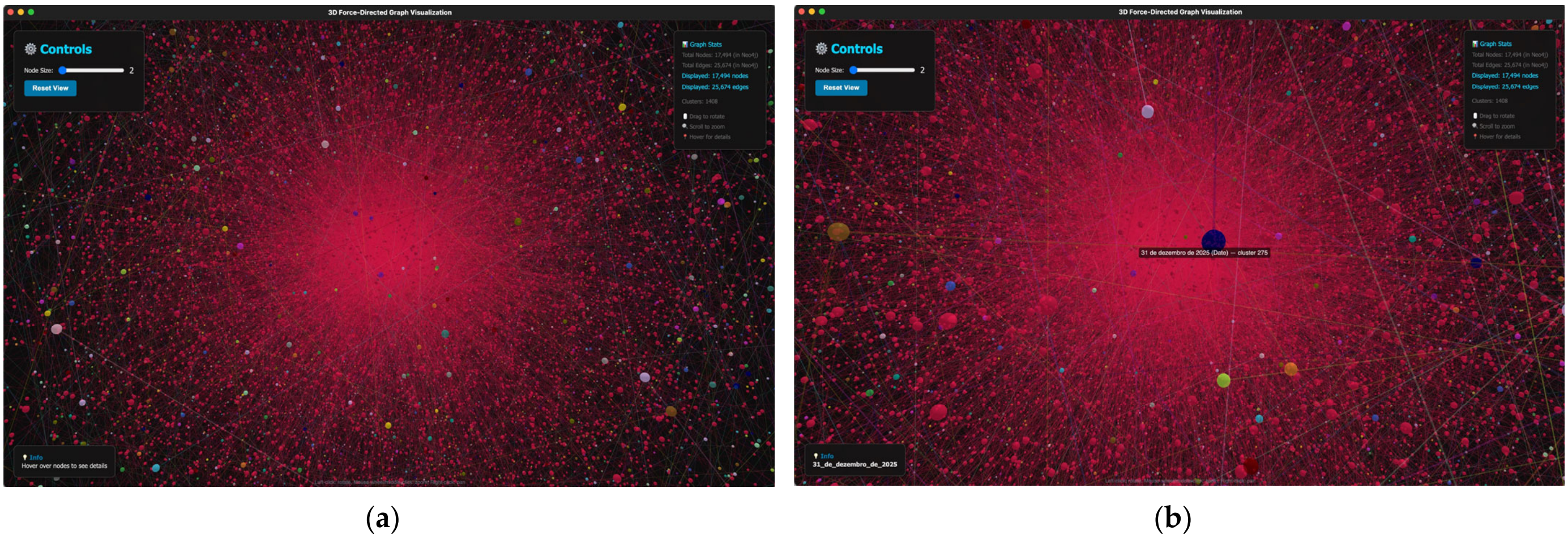

(**a**) (**b**)

**Figure 3.** Visualization of the knowledge graph after entity deduplication using 3d-force-graph. (**a**) Overview showing 17,494 entity nodes, 25,674 edges, and 1408 clusters; the dense central region and sparse periphery reflect the uneven distribution of entity connections. (**b**) Detail view showing a selected entity node ("31 de dezembro de 2025", "31 December 2025" in English, typed as Date, cluster 275), illustrating the entity-level metadata available on hover.

#### 3.3.3. Computational Cost

Another practical challenge concerns the computational cost of graph construction. Despite running the LLM-based entity extraction on an NVIDIA DGX Spark workstation (NVIDIA, Santa Clara, CA, USA) with parallel processing, the full corpus required more than 10 h on average to process. This makes iterative experimentation that includes testing different extraction prompts, merging thresholds, or graph structures a very time-consuming task. In a research project with constrained timelines, this cost effectively limited the number of optimization cycles we could explore.

#### 3.3.4. Summary of Knowledge Graph Findings

Due to the project's time constraints and the results described above, the current system deployment does not use the knowledge graph.

## 4. Discussion

This section elaborates on the findings and their broader implications, addresses this study's limitations, and relates the results to the existing literature.

### *4.1. Data Preparation as the Dominant Factor*

The most prominent outcome of the benchmark is the substantial variation in accuracy across configurations. Comparing automated frameworks at a fixed retrieval depth of K = 200, the span between the lowest-performing pipeline (DeepSeek OCR at 79.1%) and the highest one (Docling with hierarchical splitting and image descriptions at 94.1%) is approximately 15 percentage points. All of these configurations used the same LLM, the same embedding model, the same maximum chunk size, and the same retrieval parameters. The variation results solely from data preparation choices made before the text entered the vector store. This range of accuracy values suggests that in our experimental setting, the preprocessing pipeline accounts for more variation than most practitioners would expect from swapping the LLM itself.

This finding reveals another side of RAG research that has been overlooked. A considerable amount of effort in the RAG community has been put into optimizing retrieval algorithms, reranking strategies, prompt engineering, and model selection [2,3]. Our results suggest that for corpora with complex document structures, investing in the preprocessing pipeline yields a higher return. The 7.9 percentage-point gap between naïve PDFLoader (86.2%) and the best automated configuration (Docling at 94.1%) demonstrates that careful framework selection combined with appropriate splitting and metadata enrichment can achieve accuracy levels that exceed even manual curation (91.3% at K = 200).

The PDFLoader baseline registered 86.2% accuracy, higher than we initially expected. We attribute this to the number of documents in the corpus. Although it encompasses around half a million words, we believe that the naïve loader still manages to extract enough readable text for the LLM to produce reasonable answers in many cases. A larger corpus may corrode such a high level attained.

The observation that PDFLoader outperforms both Docling and MinerU when all three use recursive splitting (Section 3.2.1) merits discussion. At a high retrieval depth (K = 200) and with a basic splitter, the structural information that Docling and MinerU produce is not exploited, because the recursive splitter does not use Markdown structure or hierarchical metadata. The overhead of intermediate Markdown conversion does not pay off until a structure-aware splitter is applied. This interpretation is consistent with the large gain observed when Docling is combined with hierarchical recursive splitting (Section 3.2.3). The advantage of Docling lies not in better text extraction per se but in its ability to produce well-structured Markdown that downstream splitting and metadata enrichment can exploit, a major conclusion of the present study.

Unexpectedly, the MinerU Pipeline without transformations (config. 2: 82.8 ± 1.3%) underperformed the assumed bottom baseline, PDFLoader (config. 1: 86.2 ± 0.9%). We believe this is explained by MinerU's default output format. It generates HTML tables and preserves LaTeX formulae, which, if not converted to Markdown by a recursive splitter, introduce noise that interferes with chunking and retrieval. This makes sense, given that 94.4% of the documents (34 out of 36) in our corpus contain complex tabular content and MinerU outputs tables as HTML, which the Markdown-based splitters do not handle well.

The MinerU HTTP-client VLM variant, despite using a more capable Vision–Language Model, did not consistently improve on the Pipeline version (Section 3.2.5). Combined with its instability (crashing on 12 of 36 documents), higher computational requirements, and misrecognition of Portuguese-specific characters, the VLM variant proved impractical for our use case.

Our work is complementary to other hallucination mitigation strategies. Recent approaches operate downstream of the data pipeline: self-consistency decoding [27], chain-of-verification prompting [28], and retrieval-augmented fine-tuning [29] all aim to reduce hallucinations at the generation stage. By contrast, our focus is upstream: ensuring that the data fed to the retriever and the LLM are structurally accurate and complete. These approaches are not mutually exclusive. A high-quality preprocessing pipeline reduces the frequency of retrieval failures that trigger hallucinations, while downstream methods can catch residual errors. The 15-percentage-point accuracy span we observed across preprocessing configurations (Section 3.2) suggests that upstream data quality is a prerequisite for effective hallucination mitigation, regardless of which downstream strategy is employed.

The corpus characterization presented in Section 2.2 reveals substantial heterogeneity across document categories. Legal consolidations, which are text-dense (458 words/page) and contain no images, are predominantly targeted by straightforward text questions, a category where all frameworks score between 80.9% and 93.0% (Table 6). By contrast, manuals, which are image-heavy (695 unique images) and contain complex tables and screenshots, are targeted by table-dependent and system/SAP questions, categories where framework and splitting choices produce the largest accuracy differences. The per-question-type analysis (Section 3.2.6) confirms that table-dependent questions are the primary driver of the overall accuracy gap: PDFLoader, MinerU, and Docling with recursive splitting all score below 70.1% on these questions, while Docling with hierarchical splitting scores 99.8%. This 33-percentage-point gain on a single question type accounts for most of the overall improvement reported in Section 3.2.3, and is consistent with the corpus-level observation that 94.4% of documents contain tabular structures. A systematic per-category evaluation, linking individual document characteristics to per-question performance, is a natural direction for future work.

### 4.2. Metadata and Splitting Matter More than the Framework

Another finding that we consider particularly relevant for practitioners is that the choice of splitting strategy and metadata enrichment contributed more to accuracy than the choice of conversion framework alone. Docling without hierarchical metadata scored 84.7% (config. 16), while the same Docling output with hierarchical splitting and breadcrumb metadata reached 91.2% (config. 19), a gain of 6.5 percentage points. By comparison, the difference between Docling and PDFLoader, both with recursive splitting, was only 1.5 points (84.7% vs. 86.2%) and in fact favored PDFLoader.

These results are particularly informative. The 6.5-point gain from hierarchical splitting and metadata in Docling suggests that preserving the document's logical structure during chunking is at least as important as the accuracy of the text extraction itself. This sustains the claim that once the text extraction is reasonably accurate, the way the text is organized into chunks and enriched with structural context becomes the bottleneck. hierarchical recursive splitting, which maps the full document tree and prepends the section path to every chunk, gives the LLM explicit information about where each piece of text sits within the document, enabling it to filter irrelevant sections during Chain-of-Thought reasoning. We believe that this mechanism is especially important for our corpus, where many documents share overlapping terminology (e.g., military ranks, administrative procedures, and regulatory references) and the section context is necessary to disambiguate. This is one important conclusion resulting from these experiments: structure-aware splitting with hierarchical metadata is the single most effective lever for improving RAG accuracy in our setting.

The per-question-type analysis (Table 6) clarifies why hierarchical splitting produces such large overall gains. The 6.5-percentage-point improvement from recursive to hierarchical splitting (Section 3.2.3) is almost entirely driven by table-dependent questions, where accuracy jumps from 66.6% to 99.8%. On the other three question types, the gains are modest (0.8 to 1.3 pp). This concentration of the effect suggests that the hierarchical metadata enable the retriever to locate and correctly segment tabular content, which basic recursive splitting fragments in ways that destroy the table structure. For corpora with lower table density, the advantage of hierarchical splitting may be correspondingly smaller.

Adding VLM-based image descriptions on top of hierarchical splitting (config. 19 $\rightarrow$ 20) contributed a further 2.9 percentage points. This gain is substantial in context and comes at relatively low computational cost, since the image description step can be activated or deactivated through a single configuration parameter. Notably, config. 20 (94.1%) surpasses the hand-made Markdown baseline (91.3% at K = 200), demonstrating that a well-configured automated pipeline can exceed the quality of manual curation for this corpus. We attribute the reversal to the image descriptions. The hand-made Markdowns do not include textual representations of embedded images, whereas the automated pipeline does.

The number of retrieved chunks also plays a role, though its interaction with pipeline quality is nuanced. For the hand-made Markdowns, accuracy rises from 78.4% at K = 5 to 92.9% at K = 50, a gain of 14.5 percentage points, but slightly decreases to 91.3% at K = 200. The slight decrease at K = 200 suggests that retrieving a very large number of chunks may introduce noise that marginally dilutes the quality of the context provided to the LLM. For Docling with recursive splitting, the sensitivity to K is much weaker: only 2.2 points separate K = 5 (82.5%) from K = 200 (84.7%), and the difference between K = 50 and K = 5 is not statistically significant. This compressed range suggests that without hierarchical metadata, increasing K provides diminishing returns because the chunks themselves lack the structural context needed for effective retrieval. In both cases, the effect of metadata enrichment (6.5 pp for Docling) far exceeds the effect of increasing K.

### *4.3. Font-Based vs. LLM-Based Hierarchy Rebuilding*

In our expanded evaluation with 50 independent runs and K = 200, LLM-based hierarchy rebuilding (HR-LLM) outperformed font-based rebuilding (HR-F) in both configuration pairs tested: individually (config. 7 vs. 6: +1.0 pp, d = 0.78) and in combination with other transformations (config. 9 vs. 8: +2.4 pp, d = 1.73). Both differences are statistically significant.

The underperformance of config. 8 (HTML + LaTeX + HR-F), which is the worst-performing MinerU configuration in the benchmark, is noteworthy. It suggests that stacking font-based hierarchy rebuilding with other cleaning transformations introduces interactions that degrade output quality. By contrast, the LLM-based approach appears more robust when combined with additional processing steps.

We should note, however, that the overall differences between HR-F and HR-LLM are modest in absolute terms (1.0 to 2.4 percentage points) and that neither approach approaches the accuracy achievable through Docling's native hierarchical extraction (config. 19: 91.2%). For practitioners working with MinerU, the choice between HR-F and HR-LLM appears to be context-dependent rather than clear-cut. For our corpus of formally structured administrative documents, the results favor HR-LLM, though we cannot generalize this finding to other document types.

### 4.4. The Knowledge Graph Has Not Helped, Yet

The GraphRAG exploration produced results that, although disappointing, we believe are useful to disclose. The basic RAG pipeline with Docling achieved 94.1%, while GraphRAG scored 82%. Entity deduplication made it even slightly worse (81%). These results run counter to the expectation, supported by the recent literature [6,18,30]. Knowledge graphs should have enhanced RAG by capturing relationships that vector-based retrieval alone cannot represent. There may be several possible explanations.

First, the graph was shallow and sparse, with over 20,000 entities, but had a low average degree, meaning that most entities had few connections. According to Hossain and Sarıyüce [30], the average degree of our graph suggests that it is shallow and sparse. This kind of graph structure is not favorable for community detection algorithms such as Leiden, which GraphRAG implementations typically rely on.

Second, the LLM-based entity extraction (via LLMGraphTransformer) was not guided by a domain ontology. Entities were extracted generically, producing a graph where noise and redundancy diluted the useful structural information.

Third, the entity deduplication pipeline, while reducing the entity count from 20,055 to 17,494, appears to have merged some semantically distinct concepts (entities with surface similarity above 85% but different meanings in context), which degraded retrieval accuracy rather than improving it.

We are convinced that the merging threshold, set at 85% similarity, was too aggressive for certain entity pairs, collapsing semantically distinct concepts into a single node. This is an area that requires further refinement, possibly with domain-specific entity typing or a more conservative merging strategy.

Our GraphRAG exploration was constrained by practical factors. The 10+ hour processing time for graph construction on DGX Spark made iterative experimentation difficult. We believe that with more time and computational budget, it would be possible to refine the extraction prompts, implement domain-specific entity typing, adjust the merging threshold, and experiment with graph densification techniques. Our negative result should therefore be interpreted not as evidence that GraphRAG cannot work for this type of corpus but rather that basic, unguided graph construction is insufficient and that a well-configured basic RAG pipeline sets a high bar that GraphRAG must clear to justify its additional complexity and cost.

We report these findings because they illustrate an important practical point: basic GraphRAG, based on automated LLM entity extraction without careful ontology design or graph densification, does not necessarily outperform a well-configured basic RAG pipeline. In the present case, investing effort in data preparation and chunking strategy (achieving 94.1% with Docling) proved far more effective than adding a knowledge graph layer on top of a less optimized pipeline (82% with GraphRAG).

### 4.5. Implications for Non-English Document Processing

Our experience with Portuguese documents revealed specific challenges that are likely shared by other non-English languages. The misrecognition of the letter “ç” by MinerU’s VLM variant is not merely a secondary issue. It corrupts tokens that participate in retrieval, meaning that queries about “promoção” (promotion) might not match chunks where the word was incorrectly extracted as “promoçao”, or “promocão”, or even “promocao”. We are convinced that this type of error stems from the training data distribution of the underlying models, which are predominantly trained on English and, to a lesser extent, Chinese text.

Docling handled Portuguese characters without noticeable errors, since it has a vast selection of pre-trained OCR (we used EasyOCR), which contributed to its superior performance in our benchmark. For practitioners working with non-English corpora, particularly in languages with diacritics, cedillas, or other special characters, this is a relevant selection criterion that is not captured by existing parsing benchmarks, which are overwhelmingly English-focused [9,13,14].

*4.6. Limitations*

The present study experiences several limitations that should be considered when interpreting the results.

First, the corpus contains 36 documents. Although these span 1706 pages and nearly 492,000 words, a non-trivial volume, the number of distinct documents is limited. It is possible that the relative performance of the frameworks would change with a larger or more diverse corpus. In particular, the surprisingly strong performance of PDFLoader (86.2%) may partly be an artifact of the limited document count, where the naïve loader manages to extract sufficient text for many questions simply because the LLM has fewer documents to confuse. The system architecture, based on the Medallion Architecture with containerized deployment, is designed to support larger corpora through incremental ingestion and modular processing. However, the computational cost of certain pipeline stages (notably VLM-based image description and LLM-based hierarchy rebuilding) may become a bottleneck at scale, and the interaction between corpus size and retrieval depth (K) would need to be reassessed.

Second, our evaluation dataset comprises 50 questions. While these were carefully constructed to target specific failure modes (e.g., table extraction, hierarchy-dependent answers, and image-dependent content), a larger and more varied question set would provide stronger evidence of generalizability. We partially mitigate this limitation through the 50-run evaluation protocol, which enables statistical comparison across configurations with narrow confidence intervals. Wilcoxon signed-rank tests confirm that the key differences reported herein are statistically significant, with large effect sizes.

Third, all experiments used a single LLM (gpt-4o-mini) and a single embedding model (text-embedding-3-small). The framework rankings we observed might not hold with different models. For instance, a more capable LLM might compensate for some preprocessing errors, narrowing the gap between configurations. However, the pipeline is architecturally model-agnostic. It produces Markdown files and metadata that can be consumed by any LLM and any embedding model without modification. Because the accuracy differences we report stem from the quality of the input data rather than from model-specific interactions, we expect the relative ranking of preprocessing pipelines to hold across models, though the magnitude of the differences may vary.

Fourth, the accuracy metric relies on LLM-as-judge evaluation, which introduces its own biases and volatility. While averaging accuracy across 50 independent runs can mitigate stochastic variation and yield narrow confidence intervals, this metric may not be as rigorous as human expert evaluation.

Fifth, the maximum chunk size (1000 characters) and overlap (200 characters) were held constant across all configurations. For the recursive and Markdown recursive strategies, these values determine the chunk boundaries directly. For the hierarchical recursive strategy, chunks are additionally bounded by section markers and may therefore be shorter than the configured maximum. We did not systematically optimize these parameters. A factorial analysis of chunk size, overlap, and splitting strategy would be informative but would multiply the number of configurations beyond what was feasible in this study.

Sixth, the GraphRAG exploration was preliminary. We tested a single graph construction approach (LLMGraphTransformer with generic entity extraction) and a single deduplication strategy. A more thorough study would explore ontology-guided extrac-tion, alternative graph databases, different community detection algorithms, and hybrid retrieval strategies combining vector search with graph traversal.

Finally, the hand-made Markdowns used as the upper baseline were produced by members of the research team, which may introduce a form of researcher bias. The corrections may have been unconsciously aligned with the benchmark questions. We attempted to mitigate this by creating the Markdowns before finalizing the evaluation dataset, but the risk cannot be entirely excluded.

### 4.7. Comparison with Related Work

Existing benchmarks for PDF conversion tools, such as OmniDocBench [9] and the Docling technical report [7], focus on parsing-level metrics, text edit distance, table structure similarity (TEDS), and layout detection accuracy, and do not evaluate downstream task performance. The present study complements these benchmarks by providing a task-oriented perspective: how well does each framework support question answering when embedded in a full RAG pipeline?

Li et al. [4] recently proposed a domain-adapted RAG data pipeline for building documents, reporting approximately 30% improvement in coverage and structural preservation through tailored preprocessing. Their findings align with ours in emphasizing that domain-adapted data preparation is crucial, though their evaluation focused on information extraction metrics rather than question-answering accuracy, as in the present case.

In the GraphRAG space, Ali et al. [18] demonstrated that ontology-grounded knowledge graphs can mitigate hallucinations in clinical question answering. Their approach differs from ours in that they used pre-defined medical ontology to guide graph construction, whereas our entity extraction was fully automated and domain-agnostic. The contrast in outcomes reinforces the importance of ontological guidance for effective GraphRAG.

## 5. Conclusions

In this article, we presented a systematic evaluation of PDF conversion frameworks for RAG, measuring their impact on downstream question-answering accuracy, rather than on parsing-level metrics. We tested 21 pipeline configurations over a corpus of about half a million words in 36 administrative documents in Portuguese, using four conversion frameworks (Docling, MinerU, PDFLoader, and DeepSeek OCR) with various combinations of cleaning transformations, splitting strategies, and metadata enrichment.

The main conclusions are as follows:

- Data preparation quality is the most influential factor in RAG accuracy in our experimental setting; at a fixed retrieval depth of K = 200, the span between the worst and best automated configurations is approximately 15 percentage points, and all pairwise differences are statistically significant (Wilcoxon signed-rank tests, $p < 0.05$).
- Docling, combined with hierarchical structure extraction and VLM-based image descriptions, achieved $94.1 \pm 1.6\%$ accuracy, surpassing even the manually curated upper baseline ($91.3 \pm 1.2\%$ at K = 200); this demonstrates that a well-configured automated pipeline can match and exceed the quality of manual curation for structured administrative documents.
- The comparison between font-based (HR-F) and LLM-based (HR-LLM) hierarchy rebuilding, tested on MinerU output, yielded mixed results: HR-LLM slightly outperformed HR-F in our evaluation, though the differences are small and may be context-dependent; neither approach, however, approaches the accuracy of Docling's native hierarchical extraction.
- Additionally, our results highlight the importance of language-specific validation when applying these frameworks to non-English corpora. DeepSeek OCR exhibited systematic misrecognition of Portuguese diacritical characters (e.g., 'ç'), while Docling handled Portuguese text without noticeable errors, suggesting that framework selection should account for language-specific robustness.
- A basic GraphRAG implementation, without domain-specific ontology guidance, underperformed basic RAG by a wide margin (82% vs. 94.1%).
- We believe these findings carry a practical message for practitioners and teams building RAG systems over document collections: before optimizing retrieval or generation, invest in the data preparation pipeline. The configuration-driven architecture we propose and transparent framework swapping provide a starting point for systematic experimentation with different preprocessing strategies.

Regarding future work, we intend to expand the evaluation in several directions.

First, we plan to increase the size of both the corpus and the evaluation dataset, including documents not only from military human resources but also from other sources.

Second, we aim to evaluate the pipeline with additional LLMs, including open-source models running locally, to assess whether the framework rankings are model-dependent.

Third, we plan to apply established end-to-end RAG evaluation frameworks such as RAGAs [24] for a more comprehensive assessment.

Fourth, we intend to revisit the GraphRAG approach with domain-specific ontology design, guided entity extraction, and graph densification techniques to determine whether a more carefully constructed knowledge graph can surpass the basic RAG results we achieved.

Finally, we plan to systematically investigate the impact of chunk size and overlap parameters, which were not varied in the present study.

**Author Contributions:** Conceptualization, R.H.P., A.S., B.M.F., H.L.-C., J.D., J.L.R., L.P.R., P.P. and J.P.M.d.S.; data curation, J.G.M.d.S., R.Y., R.H.P. and A.S.; formal analysis, J.G.M.d.S., R.Y. and R.H.P.; funding acquisition, A.S. and J.P.M.d.S.; investigation, J.G.M.d.S., R.Y., R.H.P., A.S. and J.P.M.d.S.; methodology, J.G.M.d.S., R.Y., R.H.P., A.S., B.M.F., H.L.-C., J.D., J.L.R., L.P.R., P.P. and J.P.M.d.S.; project administration, A.S. and J.P.M.d.S.; resources, A.S. and J.P.M.d.S.; software, J.G.M.d.S., R.Y. and R.H.P.; supervision, R.H.P., A.S., B.M.F., H.L.-C., J.D., J.L.R., L.P.R., P.P. and J.P.M.d.S.; validation, R.H.P., A.S. and J.P.M.d.S.; visualization, J.G.M.d.S., R.Y., R.H.P. and J.P.M.d.S.; writing—original draft preparation, J.G.M.d.S., R.Y., R.H.P. and J.P.M.d.S.; writing—review and editing, R.H.P., A.S., B.M.F., H.L.-C., J.D., J.L.R., L.P.R., P.P. and J.P.M.d.S. All authors have read and agreed to the published version of the manuscript.

**Funding:** This research study was funded by Fundação para a Ciência e a Tecnologia (FCT), grant number 2024.07619.IACDC. LIACC affiliates are financially supported by UID/00027/2025 (LIACC—Artificial Intelligence and Computer Science Laboratory; DOI https://doi.org/10.54499/UID/00027/2025), funded by Fundação para a Ciência e a Tecnologia, I.P./MECI, through national funds.

**Institutional Review Board Statement:** Not applicable.

**Informed Consent Statement:** Not applicable.

**Data Availability Statement:** The source code and pipeline configurations are available at https://github.com/sousaalexandre/loss-j. The document corpus used in this study consists of publicly accessible documents from the Portuguese Army and can be obtained upon request from the corresponding author. The evaluation dataset (50 questions with expected answers) is available in the project repository.

**Acknowledgments:** The authors acknowledge the Personnel Command of the Portuguese Army for providing the document corpus used in this study. During the preparation of this manuscript, the authors used Claude (Anthropic, San Francisco, CA, USA, Claude Opus 4.6) to draft initial versions of certain sections and to structure the manuscript. The authors have reviewed and edited all AI-generated output and take full responsibility for the content of this publication.

**Conflicts of Interest:** The authors declare no conflicts of interest. The funders had no role in the design of the study; in the collection, analyses, or interpretation of data; in the writing of the manuscript; or in the decision to publish the results.

## Abbreviations

The following abbreviations are used in this manuscript:

| | |
|---|---|
| API | Application Programming Interface |
| BAAI | Beijing Academy of Artificial Intelligence |
| ETL | Extract, Transform, Load |
| HR-F | Hierarchy Rebuilding—Font-based |
| HR-LLM | Hierarchy Rebuilding—LLM-based |
| HTML | HyperText Markup Language |
| JSON | JavaScript Object Notation |
| K | Number of Retrieved Top-k chunks |
| LLM | Large Language Model |
| OCR | Optical Character Recognition |
| PDF | Portable Document Format |
| RAG | Retrieval-Augmented Generation |
| REST | Representational State Transfer |
| TEDS | Tree Edit Distance Similarity |
| VLM | Vision–Language Model |

## Appendix A

The appendix presents the translated and original prompts used to assess responses during the LLM-as-judge process. This prompt was originally written in European Portuguese, consistent with the corpus language and the expected answers. The English translation is provided here for readability.

### *Appendix A.1*

Translation to English:

prompt_template = """

### Evaluation Task (European Portuguese)

You are an expert evaluator of a RAG (Retrieval-Augmented Generation) system. Your task is to evaluate the 'RAG Response' based on the user's 'Query', using the 'Expected Response' as the ground truth or ideal answer.

Compare the 'RAG Response' with the 'Expected Response' based on the following criteria:

1. **Factual Accuracy:** Does the 'RAG Response' contain factually correct information, as defined by the 'Expected Response'?

2. **Completeness:** Does the 'RAG Response' cover all key points and the central meaning present in the 'Expected Response'?

### Scoring Criteria (0–100)

* **95–100 (Perfect):** The 'RAG Response' perfectly answers the 'Query' and is 100% accurate and complete. It contains all key information from the 'Expected Response'. Differences in style are acceptable.

* **80–94 (Good):** The 'RAG Response' is factually accurate and answers the 'Query' well but may omit minor details or be slightly less complete than the 'Expected Response'.

* **50–79 (Fair):** The 'RAG Response' answers the 'Query' but is noticeably incomplete or omits important facts present in the 'Expected Response'.

* **10–49 (Poor):** The 'RAG Response' contains factual inaccuracies or "hallucinations" (information that contradicts the 'Expected Response') or is largely irrelevant to the 'Query'.

* **0–9 (Terrible):** Completely wrong or irrelevant.

Focus solely on factual accuracy and completeness. Do not penalize differences in style or wording, **provided that** the central meaning is the same.

---

### Data for Evaluation

**Query:**

"{query}"

**Expected Response (Ideal):**

"{expected}"

**RAG Response (Received):**

"{received}"

"""

*Appendix A.2*

Original in European Portuguese:

prompt_template = """

### Tarefa de Avaliação (Português Europeu)

És um avaliador especialista de um sistema RAG (Retrieval-Augmented Generation).

A tua tarefa é avaliar a 'Resposta RAG' com base na 'Query' do utilizador, usando a 'Resposta Esperada' como a fonte de "verdade" ou a resposta ideal.

Compara a 'Resposta RAG' com a 'Resposta Esperada' com base nos seguintes critérios:

1. **Precisão Factual:** A 'Resposta RAG' contém informação factualmente correta, tal como definido pela 'Resposta Esperada'?

2. **Completude:** A 'Resposta RAG' cobre todos os pontos-chave e o significado central presentes na 'Resposta Esperada'?

### Critérios de Pontuação (0–100)

* **95–100 (Perfeita):** A 'Resposta RAG' responde perfeitamente à 'Query' e é 100% precisa e completa. Contém toda a informação chave da 'Resposta Esperada'. Diferenças de estilo são aceitáveis.

* **80–94 (Boa):** A 'Resposta RAG' é factualmente precisa e responde bem à 'Query', mas pode omitir pequenos detalhes ou ser ligeiramente menos completa que a 'Resposta Esperada'.

* **50–79 (Razoável):** A 'Resposta RAG' responde à 'Query', mas é visivelmente incompleta ou omite factos importantes presentes na 'Resposta Esperada'.

* **10–49 (Fraca):** A 'Resposta RAG' contém imprecisões factuais, "alucinações" (informação que contradiz a 'Resposta Esperada') ou é maioritariamente irrelevante para a 'Query'.

* **0 (Terrível):** Completamente errada ou irrelevante.

Foca-te apenas na precisão factual e na completude. Não penalizes por diferenças de estilo ou formulação, *desde que* o significado central seja o mesmo.

---

### Dados para Avaliar

**Query:**

"{query}"

**Resposta Esperada (Ideal):**

"{expected}"

**Resposta RAG (Recebida):**

"{received}"

"""

## Appendix B

**Table A1.** Shapiro–Wilk normality test results for the paired differences across all 24 pairwise comparisons. The normality assumption is rejected at $\alpha = 0.05$ in 18 of 24 cases, motivating the use of the Wilcoxon signed-rank test as the primary statistical test (Section 2.6).

| Comparison | Configurations | | W | $p$ | Normal? |
|---|---|---|---|---|---|
| PDFLoader vs. MinerU (Recursive) | 1 | 2 | 0.5988 | $1.9 \times 10^{-10}$ | No |
| PDFLoader vs. Docling (Recursive) | 1 | 16 | 0.7577 | $1.1 \times 10^{-7}$ | No |
| Docling vs. MinerU (Recursive) | 16 | 2 | 0.9888 | 0.914 | Yes |
| MinerU: Recursive vs. Md Recursive | 2 | 3 | 0.9345 | 0.008 | No |
| MinerU: base vs. HTML | 3 | 4 | 0.6908 | $5.7 \times 10^{-9}$ | No |
| MinerU: base vs. LaTeX | 3 | 5 | 0.7275 | $2.7 \times 10^{-8}$ | No |
| MinerU: base vs. HR-F | 3 | 6 | 0.8561 | $2.3 \times 10^{-5}$ | No |
| MinerU: base vs. HR-LLM | 3 | 7 | 0.8363 | $6.7 \times 10^{-6}$ | No |
| MinerU: HR-F vs. HR-LLM (indiv.) | 6 | 7 | 0.9827 | 0.669 | Yes |
| MinerU: HR-F vs. HR-LLM (comb.) | 8 | 9 | 0.9863 | 0.828 | Yes |
| Docling: Recursive vs. Hierarchical | 16 | 19 | 0.9470 | 0.026 | No |
| Docling: Hierarchical vs. +Images | 19 | 20 | 0.8838 | $1.5 \times 10^{-4}$ | No |
| Docling: Recursive vs. Hier. + Images | 16 | 20 | 0.9395 | 0.013 | No |
| Docling + Img vs. Hand-made (K = 200) | 20 | 12 | 0.8795 | $1.1 \times 10^{-4}$ | No |
| Docling + Img vs. Hand-made (K = 50) | 20 | 13 | 0.6458 | $9.9 \times 10^{-10}$ | No |
| Hand-made: K = 50 vs. K = 200 | 13 | 12 | 0.9665 | 0.166 | Yes |
| Hand-made: K = 50 vs. K = 20 | 13 | 14 | 0.8371 | $7.1 \times 10^{-6}$ | No |
| Hand-made: K = 20 vs. K = 5 | 14 | 15 | 0.8614 | $3.2 \times 10^{-5}$ | No |
| Docling: K = 200 vs. K = 50 | 16 | 17 | 0.8976 | $4.0 \times 10^{-4}$ | No |
| Docling: K = 50 vs. K = 5 | 17 | 18 | 0.9703 | 0.237 | Yes |
| MinerU Pipeline vs. VLM (no transf.) | 3 | 10 | 0.7507 | $7.6 \times 10^{-8}$ | No |
| MinerU Pipeline vs. VLM (full clean.) | 8 | 11 | 0.9762 | 0.404 | Yes |
| Best vs. Worst (Docling vs. DeepSeek) | 20 | 21 | 0.8334 | $5.7 \times 10^{-6}$ | No |
| PDFLoader vs. Docling + Images | 1 | 20 | 0.6779 | $3.4 \times 10^{-9}$ | No |